\documentclass[aps, prb, amsmath, twocolumn, longbibliography, 10pt]{revtex4-1}

\usepackage{graphicx} 
\usepackage{multibib} 
\usepackage{xcolor}
\newcites{supp}{ } 
\usepackage{soul} 
\usepackage{multirow}

\begin{document}

\title{Extended-SWIR GeSn LEDs with reduced footprint and power consumption}

\author{Mahmoud R. M. Atalla$^{\dagger}$}
\affiliation{Department of Engineering Physics, \'Ecole Polytechnique de Montr\'eal, C.P. 6079, Succ. Centre-Ville, Montr\'eal, Qu\'ebec, Canada H3C 3A7}
\author{Youngmin Kim$^{\dagger}$}
\affiliation{School of Electrical and Electronic Engineering, Nanyang Technological University, 50 Nanyang Avenue, Singapore 639798, Singapore.}
\affiliation{$^{\dagger}$These authors contributed equally to this work.}
\author{Simone Assali}
\affiliation{Department of Engineering Physics, \'Ecole Polytechnique de Montr\'eal, C.P. 6079, Succ. Centre-Ville, Montr\'eal, Qu\'ebec, Canada H3C 3A7}
\author{Daniel Burt}
\affiliation{School of Electrical and Electronic Engineering, Nanyang Technological University, 50 Nanyang Avenue, Singapore 639798, Singapore.}
\affiliation{$^{\dagger}$These authors contributed equally to this work.}
\author{Donguk Nam*}
\affiliation{School of Electrical and Electronic Engineering, Nanyang Technological University, 50 Nanyang Avenue, Singapore 639798, Singapore.}
\affiliation{$^{\dagger}$These authors contributed equally to this work.}

\author{Oussama Moutanabbir}
\email{oussama.moutanabbir@polymtl.ca and nam@ntu.edu.sg}
\affiliation{Department of Engineering Physics, \'Ecole Polytechnique de Montr\'eal, C.P. 6079, Succ. Centre-Ville, Montr\'eal, Qu\'ebec, Canada H3C 3A7}

\begin{abstract}
$^{\dagger}$These authors contributed equally to this work.
\end{abstract}

\begin{abstract}
CMOS-compatible short- and mid-wave infrared emitters are highly coveted for the monolithic integration of silicon-based photonic and electronic integrated circuits to serve a myriad of applications in sensing and communications. In this regard, a group IV germanium-tin (GeSn) material epitaxially grown on silicon (Si) emerges as a promising platform to implement tunable infrared light emitters. Indeed, upon increasing the Sn content, the bandgap of GeSn narrows and becomes direct, making this material system suitable for developing an efficient silicon-compatible emitter. With this perspective, microbridge PIN GeSn LEDs with a small footprint of \boldmath{$1,520$ $\mu$}m$^2$ are demonstrated and their operation performance is investigated. The spectral analysis of the electroluminescence emission exhibits a peak at $2.31$ $\mu$m and it red-shifts slightly as the driving current increases. It is found that the microbridge LED operates at a dissipated power as low as $10.8$ W at room temperature and just $3$ W at $80$ K.  This demonstrated low operation power is comparable to that reported for LEDs having a significantly larger footprint reaching $10^6$ $\mu$m$^2$. The efficient thermal dissipation of the current design helped to reduce the heat-induced optical losses, thus enhancing light emission. Further performance improvements are envisioned through thermal and optical simulations of the microbridge design. The use of GeSnOI substrate for developing a similar device is expected to improve optical confinement for the realization of electrically driven GeSn lasers.
\end{abstract}

\maketitle

\noindent Reducing the footprint of CMOS-compatible optical interconnects is a key paradigm to address the relentless course toward the very dense integration of chips while maintaining high bandwidth and low power consumption \cite{Soref2014,Saito2014,Deen2012}. In fact, as future electronic chips require further device scaling, the associated miniaturization of metal interconnects suffers fundamental limits leading to high impedance and joule heating, which consequently limits the bandwidth. This stimulates a remarkable interest in merging silicon photonics with silicon-based integrated circuits for more efficient inter-chip and intra-chip communication, besides added multifunctional abilities such as on-chip optical modulation, spectroscopy, and sensing applications \cite{Soref2006}. However, the development of CMOS-compatible on-chip light sources remains a huge challenge owing to the lack of direct bandgap group IV semiconductors \cite{Liang2010,Zhou2015}. Recently, GeSn has become a promising solution to overcome this limitation since GeSn becomes a direct bandgap material when sufficient Sn content is incorporated in Ge lattice \cite{Moutanabbir2021,VondenDriesch2015}. Indeed, the indirect-direct bandgap crossover occurs at a Sn content of around $8$ at. \% in relaxed GeSn. Increasing the Sn content beyond this threshold improves the bandgap directness and lowers its energy, enhancing the light emission efficiency \cite{wirths2015lasing}. Photodetector devices  made of chemical vapor deposition (CVD) and molecular beam epitaxy (MBE) grown layers and heterostructures have been demonstrated to operate at wavelengths up to $4.6$ $\mu$m in the mid-infrared (MIR) spectral range \cite{Atalla2021}. While optically pumped GeSn-based lasers have been reported by many researchers recently \cite{Moutanabbir2022}, experimental attempts to develop the electrically injected counterparts remain relatively scarce \cite{Buca2022,Marzban2022}.

\noindent To achieve the dense integration of Si-based integrated circuits and silicon photonics, emitters with a small footprint and reduced power consumption are strongly desired. Several efforts have been expended toward this goal \cite{Oehme2011,Tseng2013,Gupta2013,Gallagher2015,Stange2017,Peng2020}. While early GeSn LEDs feature low power consumption of ~3.75 W, these devices have a remarkably large footprint area of around $10^6$ $\mu$m$^2$ \cite{Gupta2013}. Other LED devices reported later on had two orders of magnitude smaller device sizes of around $31,400$ $\mu$m$^2$, but they require more than $4$-fold higher power consumption \cite{Huang2019}. Herein, we present a GeSn microscale LED with a record-small footprint of around $1,520$ $\mu$m$^2$ yet operating at a low dissipated power around $3$ W at $80$ K and around $10.8$ W at room temperature. This is realized through a cavity LED utilized in a microbridge structure.

\bigskip




\begin{figure*}[htbp]
\centering
\includegraphics[width=\textwidth]{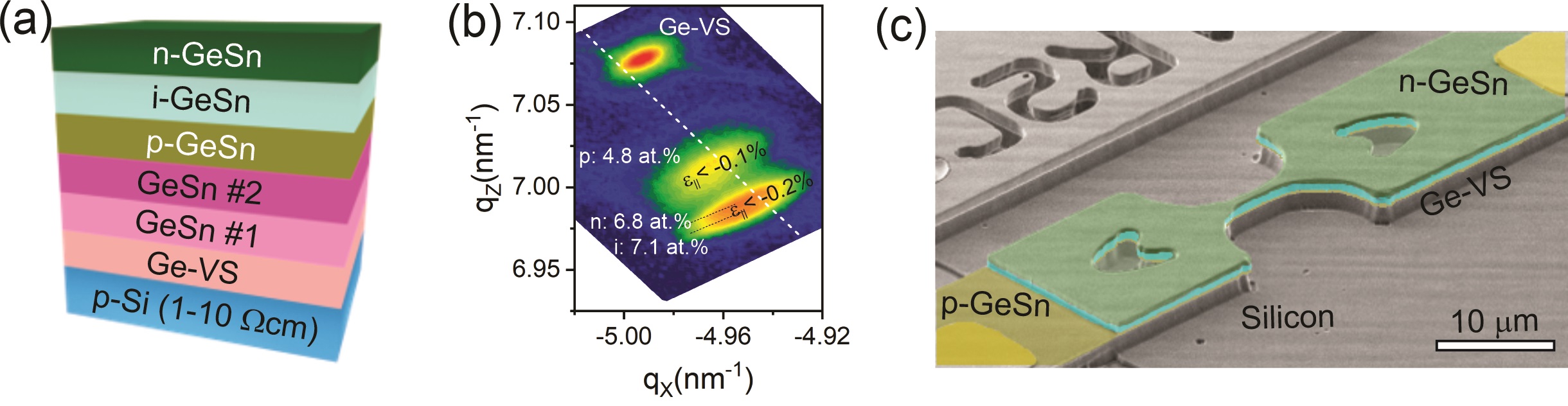}
\caption{(a) Schematic of the multilayer PIN GeSn growth on Si. (b) RSM measurement around the asymmetrical (224) XRD reflection peak indicating the Sn content and strain in the multilayer structure in panel (a). (c) Tilted SEM depicting the microbridge PIN GeSn LED. }
\label{fig:Fig__1}
\end{figure*}

\bigskip

\section{Results and discussion}
\label{sec:Results}
\noindent {\bf Growth and characterization of GeSn epilayers and LED fabrication.}

\noindent The GeSn LED samples were grown using a low-pressure CVD reactor following a multilayer growth to control lattice strain and Sn content, as schematically illustrated in Fig. 1(a). Starting from a $4$-inch Si (100) wafer, a $2.67$ $\mu$m thick Ge virtual substrate (VS) was grown using $10$\% monogermane (GeH$_{4}$) and a constant ultrapure H$_{2}$ flow implementing a two-temperature growth process at $450$ and $600$ $^\circ$C, followed by thermal cyclic annealing above $800$ $^\circ$C. Two GeSn buffer layers, with thicknesses of $140$ and $208$ nm, were grown using GeH$_{4}$ and tin-tetrachloride (SnCl$_{4}$) precursors at initial molar fractions of ($1.2$ $\times \, 10^{-2}$) and ($9.1$ $\times \, 10^{-6}$), respectively. The second GeSn buffer layer was grown at a low temperature allowing more Sn to be substitutionally incorporated into the Ge lattice. Then, PIN GeSn layers were grown at thicknesses of $375$, $830$, and $330$ nm, respectively. The p- and n-type doping were achieved using diborane (B$_{2}$H$_{6}$) and arsine (AsH$_{3}$) precursors, respectively. According to the X-ray diffraction (XRD) reciprocal space mapping (RSM) measured around the asymmetrical (224) reflection peak, as shown in Fig. 1(b), the Sn content is estimated to be $4.1$ at. \% (\#1), $6$ at. \% (\#2), $4.8$ at. \% (p-GeSn), $7.1$ at. \% (i-GeSn), $6.8$ at. \% (n-GeSn), respectively. It is also noticeable that the layers are remarkably relaxed with the highest bi-axial compressive strain below $-0.2$ \% in the i- and n-GeSn layers. 

\noindent Figure 1(c) displays a scanning electron microscope (SEM) image depicting the fabricated microbridge PIN GeSn LED device. The fabricated device has a microbridge structure for the diodes with a corner-cube cavity which was intended to enhance the light emission via optical confinement \cite{armand2019lasing}. The microbridge structure with corner-cube mirrors were patterned using photolithography, and the defined pattern was transferred onto the GeSn and Ge VS layers using reactive ion etching (RIE) with Cl$_{2}$ chemistry. A second photolithography step was performed to anisotropically etch down to expose the p-doped GeSn layer at one end of the device. Thus, the device had exposed p- and n-doped GeSn areas on both ends of the microbridge structure. The patterns for the metal contacts were then defined using photolithography at the p- and n-doped GeSn areas. Subsequently, the Cr and Au contacts with thicknesses of $20$ nm and $180$ nm, respectively, were deposited as p- and n-layers using electron beam evaporation, followed by the lift-off process.

\bigskip
\noindent {\bf GeSn PIN LED characterization and performance comparison.}

\noindent The microbridge device characterization started by measuring the I-V curves using a source meter unit over a wide bias range of up to $5$ V in the forward and reverse biases as shown in Fig. 2(a). The LED operates under forward bias reaching up to $2.16$ mA at $5$V, which corresponds to a power consumption as low as $10.8$ W at such relatively high operation bias. The large current under the reverse bias, however, indicates carrier leakage through the depletion region of the device possibly owing to various mechanisms including SRH, tunnel assisted, and diffusion leakages \cite{Son2020,Dong2017,Atalla2023}.

\bigskip

\begin{figure*}[htbp]
\centering
\includegraphics[width=\textwidth]{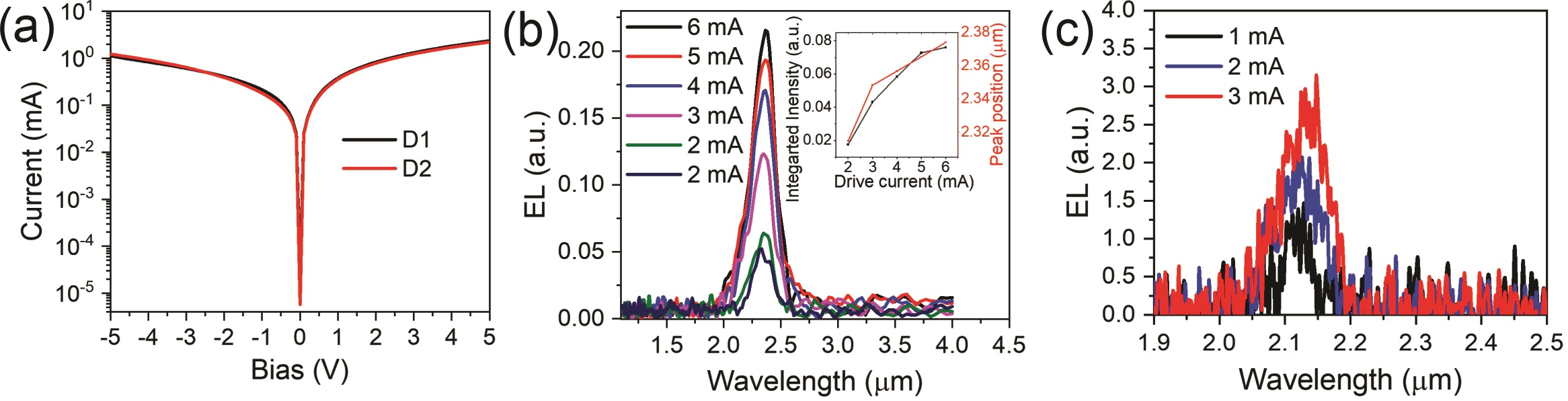}
\caption{(a) I-V measurement for two microbridge PIN LEDs showing the forward and reverse bias operation. (b) Electroluminescence (EL) as function of wavelength at the indicated LED drive current at room temperature. Inset: The integrated intensity and peak position as the drive current increases. (c) same as (b) but at 80 K temperature.}
\label{fig:Fig__1}
\end{figure*}

\bigskip

\noindent To measure and analyze the LED light emission, a Fourier transform infrared spectroscopy (FTIR) setup was utilised to measure the spectral emission from the microbridge devices. The setup mainly consists of a Bruker FTIR spectrometer, InSb liquid nitrogen cooled photodetector (PD), $40\times$ reflective objective lens, and a translation stage. To amplify the InSb PD signal, it was fed into a current preamplifier and a lock-in amplifier. The signal was re-entered into the FTIR for data processing. The lock-in technique requires the LED light to take the form of pulses either by mounting a mechanical chopper in the light path or by applying AC drive current to the LED. As shown in Fig. 2(b), the electroluminescence (EL) was measured at a driving current from $2$ mA to $6$ mA with $1$ mA stepwise. It is evident that the LED emission peak is centered around $2.31$ $\mu$m at the lowest drive current, and it red-shifts as the drive current increases, as shown in the inset of Fig. 2(b). The inset also depicts that the integrated intensity of the EL emission initially increases linearly as the driving current increases, while it saturates at $6$ V, indicating the drop in the LED emission efficiency possibly due to the heating effect. 

\noindent The device was also characterized at a cryogenic temperature of $80$ K, showing that a clear signal can be measured at a driving current as low as $1$ mA. Figure 2(c) shows the EL emission spectra measured at $80$ K. It is also clear that the EL peak blue-shifts to $2.08$ $\mu$m at $80$ K compared to a significantly longer wavelength of $2.31$ $\mu$m measured at room temperature, which is attributed to the increased bandgap at a lower temperature.

\noindent Table 1 compares the GeSn LEDs reported until today with a focus on comparing the dissipated power and the device footprint. The table clearly shows that our microbridge LED devices have the lowest combined footprint ($\mu$m$^2$) and power dissipation (I$\cdot$V) at room temperature and $80$ K. 




\begin{figure*}[htbp]
\centering
\includegraphics[width=\textwidth]{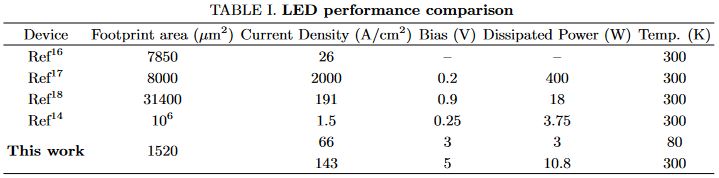}
\label{fig:Fig__1}
\end{figure*}

\bigskip

\begin{figure*}[htbp]
\centering
\includegraphics[width=\textwidth]{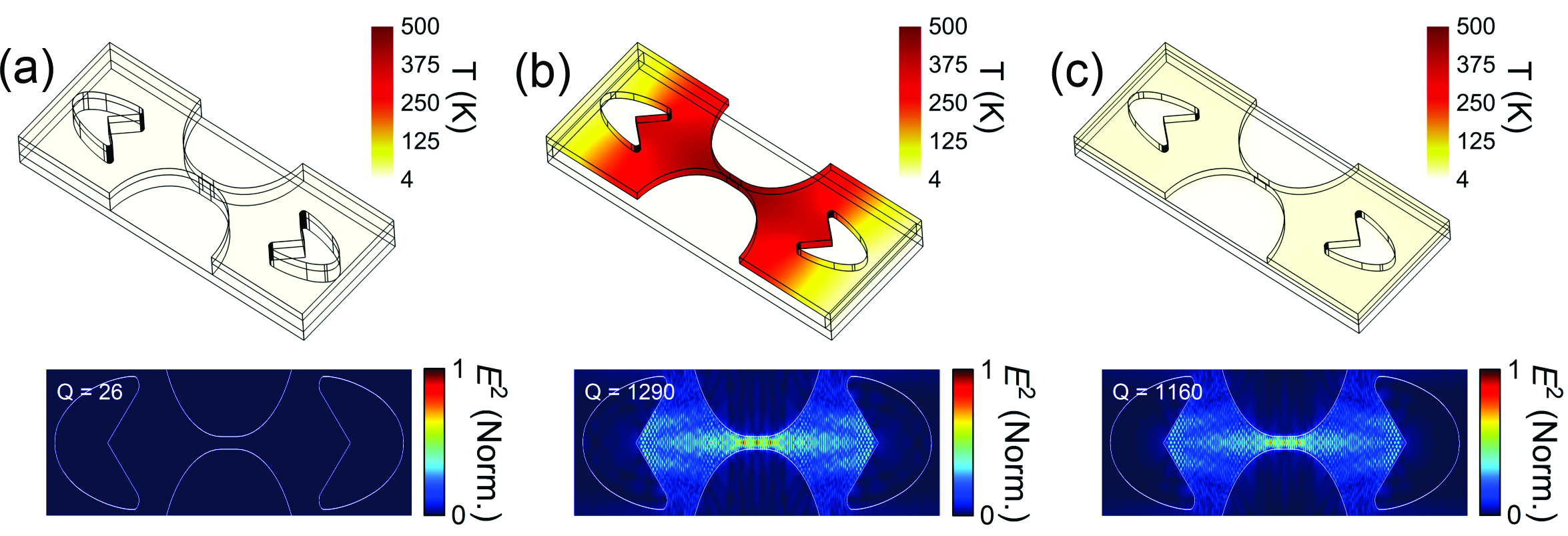}
\caption{Simulation for the thermal dissipation and optical confinement for the microbridge LED in three cases: (a) unreleased structure without under etch of Ge VS (current device). (b) Suspended structure with Ge VS under-etched. (c) Complete under etch of the Ge VS layer to fully release the PIN GeSn microbridge layer and place it on SiO2 substrate.}
\label{fig:Fig__1}
\end{figure*}

\bigskip

\noindent {\bf Corner-cube cavity light confinement effects for improving LED emission and gain properties}
\bigskip

\noindent To investigate why the cavity effect is not achieved in our LEDs and provide a viable pathway towards achieving cavity effects and possibly lasing, we performed thermal and optical simulations for the LED structures. Figure 3 presents simulated thermal (top) and optical fields (bottom) distributions calculated using a finite-element method (FEM) and finite-difference time-domain (FDTD), respectively, for three different LED structures. The three structures mimicked the structure used in our experiments and they consisted of: (i) the current device with no under etch (Fig. 3(a)), (ii) suspended structure with Ge VS layer under-etched except at the two ends of the device (Fig. 3(b)), and (iii) ideal fully released GeSn layer (Fig. 3(c)), which is possible by utilizing a GeSn-on-insulator (GeSnOI) structure \cite{Burt2021,Joo2021}. In the thermal simulation, a heat source with a current density of $300$ A\/cm$^2$ was applied along the current path. The optical field distributions for the three structures were calculated at $2155$ nm in the optical simulation. All the simulations were conducted at $4$ K. As shown in Fig. 3(a), the LED used in our experiments shows superior thermal conduction with a slightly increased temperature of $\sim  10$ K but it has poor optical confinement with a very low Q factor of $26$ due to the significant leakage of optical fields from the GeSn active layer to the Ge VS layer underneath. This explains why the cavity effect is not observed during our experiments. To address the limited optical confinement, we designed and investigated a structure suspended in the air that can be fabricated by etching the Ge VS layer \cite{Gupta2013b}. In stark contrast to poor optical confinement in the structure used in our experiments, a suspended structure, as shown in Fig. 3(b), shows strong optical confinement with a high Q factor of $1290$ because large refractive index difference between the GeSn active layer and air prevents the leakage of optical fields. The temperature of an active area, however, is increased significantly to $\sim 500$ K upon the current injection due to poor thermal dissipation in the suspended device geometry. As the increased temperature hinders optical emission in the active area due to heat-induced optical loss and also due to the reduced optical gain at an elevated temperature \cite{Kim2022}, an optimized structure was envisioned to achieve both good optical confinement and thermal dissipation. Fig. 3(c) presents an optimized structure in which a GeSn active layer is laid down onto a SiO$_2$ substrate. It is worth mentioning that the optimized structure can be fabricated using the GeSnOI substrate, which has recently been successfully demonstrated by several research groups \cite{Moutanabbir2022,Burt2021,Joo2021,Wang2021}. The optimized structure shows the simultaneous achievement of good heat dissipation with slightly increased temperature of $\sim 20$ K and superior optical confinement with a high Q factor of $1160$. We believe that it is feasible to achieve lasing in the optimized structure.

\section{Conclusion}
In summary, microbridge PIN GeSn LEDs were demonstrated and their operation performance was assessed using electrical and optical characterizations. It was found that the microbridge LED has power dissipation as low as $3$ W and a small footprint of $1,520$ $\mu$m$^2$. The efficient thermal dissipation of the current design helped reduce the heat-induced optical losses and thus enhanced light emission. According to thermal and optical simulations, the microbridge design can be promising for lasing if a similar device was fabricated on GeSnOI substrate.  

\bigskip

\noindent {\bf Acknowledgements}.
The work carried out in Montréal was supported by Natural Science and Engineering Research Council of Canada (Discovery, SPG, and CRD Grants), Canada Research Chairs, Canada Foundation for Innovation, Mitacs, PRIMA Québec, and Defence Canada (Innovation for Defence Excellence and Security, IDEaS). The research performed at Nanyang Technological University was supported by Ministry of Education, Singapore, under grant AcRF TIER 1 2019-T1-002-050 (RG 148/19 (S)). The research of the project was also supported by Ministry of Education, Singapore, under grant AcRF TIER 2 (MOE2018-T2-2-011 (S)). This work was also supported by National Research Foundation of Singapore through the Competitive Research Program (NRF-CRP19-2017-01). This work was also supported by National Research Foundation of Singapore through the NRF-ANR Joint Grant (NRF2018-NRF-ANR009 TIGER). This work was also supported by the iGrant of Singapore A*STAR AME IRG (A2083c0053). The authors would like to acknowledge and thank the Nanyang NanoFabrication Centre (N2FC).

\medskip

\medskip
\noindent {\bf Author information}.
Mahmoud R. M. Atalla and Youngmin Kim have cotributed equally to this work.
Correspondence and requests for materials should be addressed to~:\\ oussama.moutanabbir@polymtl.ca 
and dnam@ntu.edu.sg

\medskip
\noindent {\bf Data availability}.
The data that support the findings of this study are available from the corresponding author upon reasonable request.

\bibliographystyle{naturemag}
\bibliography{main}


\end{document}